\documentclass{emulateapj}

\newcommand{\avgn}{\langle N \rangle}
\newcommand{\avgm}{\langle \mu \rangle}
\newcommand{\avgx}{\langle x_H \rangle}

\begin{document}

\title{Co-orbital Oligarchy}
\author{Benjamin F. Collins\altaffilmark{1} and Re'em Sari\altaffilmark{1,2}}
\altaffiltext{1}{California Institute of Technology, MC 130-33,
        Pasadena, CA 91125}
\altaffiltext{2}{Racah Institute of Physics, Hebrew University, Jerusalem 91904,
Israel}
\email{bfc@tapir.caltech.edu}

\begin{abstract}
We present a systematic examination of the changes in semi-major axis
caused by the mutual interactions of a group of massive bodies
orbiting a central star in the presence of eccentricity dissipation.
For parameters relevant to the oligarchic stage of planet formation,
dynamical friction keeps the typical eccentricities small and prevents
orbit crossing.  Interactions at impact parameters greater than
several Hill radii cause the protoplanets to repel each other; if the
impact parameter is instead much less than the Hill radius, the
protoplanets shift slightly in semi-major axis but remain otherwise
unperturbed.  If the orbits of two or more protoplanets are separated
by less than a Hill radius, they are each pushed towards an
equilibrium spacing between their neighbors and can exist as a stable
co-orbital system.  In the shear-dominated oligarchic phase of planet
formation we show that the feeding zones contain several oligarchs
instead of only one.  Growth of the protoplanets in the oligarchic
phase drives the disk to an equilibrium configuration that depends on
the mass ratio of protoplanets to planetesimals, $\Sigma/\sigma$.
Early in the oligarchic phase, when $\Sigma/\sigma$ is low, the
spacing between rows of co-orbital oligarchs are about 5 Hill radii wide, rather
than the $10$ Hill radii cited in the literature.  It is likely that
at the end of oligarchy the average number of co-orbital oligarchs is
greater than unity.  In the outer solar system this raises the disk mass
required to form the ice giants.  In the inner solar system this
lowers the mass of the final oligarchs and requires more giant impacts
than previously estimated.  This result provides additional evidence
that Mars is not an untouched leftover from the oligarchic phase, but 
must be composed of several oligarchs assembled through giant impacts.
\end{abstract}


\section{Introduction}

The early stages in the formation of planetary systems are well
described by statistical calculations of the evolution of mass
distributions and velocity dispersions.  As larger bodies accumulate
from the swarm of proto-planetary material, their individual dynamics
begin to dominate their evolution.  \citet{Lis87} pointed out that the
finite cross-section for accretion limits the growth of each
protoplanet.  This is now known as the ``oligarchic phase.''
\citep{KI98}.  Numerical \citep{KB05,FC07,LM07} and analytical
\citep{GLSfinalstages} work has explored the transition from
oligarchic growth to the chaotic final assembly of the planets.  In
this work we examine the interactions of a moderate number of
protoplanets in an oligarchic configuration and find that neighboring
protoplanets stabilize co-orbital systems of two or more protoplanets.
We present a new picture of oligarchy in which each part of the disk
is not ruled by one but by several protoplanets having almost the same
semi-major axis.

Our approach is to systematize the interactions between each pair of
protoplanets in a disk where a swarm of small icy or rocky bodies, the
planetesimals, contain most of the mass.  The planetesimals provide
dynamical friction that circularizes the orbits of the protoplanets.
The total mass in planetesimals at this stage is more than that in
protoplanets so dynamical friction balances the excitations of
protoplanets' eccentricities.  We characterize the orbital evolution
of a protoplanet as a sequence of interactions occurring each time it
experiences a conjunction with another protoplanet.  The number
density of protoplanets is low enough that it is safe to neglect
interactions between three or more protoplanets.

To confirm our description of the dynamics and explore its application
to more realistic proto-planetary situations we perform many numerical
N-body integrations.  We use an algorithm optimized for mostly
circular orbits around a massive central body.  As integration
variables we choose six constants of the motion of an unperturbed
Keplerian orbit.  As the interactions between the other bodies in the
simulations are typically weak compared to the central force, the
variables evolve slowly.  We employ a 4th-order Runge-Kutta
integration algorithm with adaptive time-steps
\citep{numericalrecipies} to integrate the differential equations.
During periods of little interaction, the slow evolution of our
variables permits large time-steps.

During a close encounter, the inter-particle gravitational attraction
becomes comparable to the force from the central star.  In the limit
that the mutual force between a pair of particles is much stronger than the
central force, the motion can be more efficiently described as a
perturbation of the two-body orbital solution of the bodies around
each other.  We choose two new sets of variables: one to describe the
orbit of the center-of-mass of the pair around the central star, and
another for relative motion of the two interacting objects.  These
variables are evolved under the influence of the remaining particles
and the central force from the star.

Dynamical friction, when present in the simulations, is included with
an analytic term that damps the eccentricities and inclinations of
each body with a specified timescale.  All of the simulations
described in this work were performed on Caltech's Division of
Geological and Planetary Sciences Dell cluster.

We review of some basic results from the three-body problem in section
\ref{secDamped3Body}, and describe the modifications of these results
due to eccentricity dissipation.  In section \ref{secDampedNBody}, we
generalize the results of the three-body case to an arbitrary number
of bodies, and show the resulting formation and stability of
co-orbital sub-systems.  Section \ref{secOligarchy} demonstrates that
an oligarchic configuration with no initial co-orbital systems can
acquire such systems as the oligarchs grow.  Section \ref{secNEq}
describes our investigation into the properties of a co-orbital
oligarchy, and section \ref{secIsolation}
places these results in the context of the final stages of planet
formation.  The conclusions are summarized in section
\ref{secConclusions}.

\section{The Three-Body Problem}
\label{secDamped3Body}
The circular restricted planar three-body problem refers to a system
of a zero mass test particle and two massive particles on a circular
orbit.  We call the most massive object the star and the other the
protoplanet.  The mass ratio of the protoplanet to the star is $\mu$.
Their orbit has a semi-major axis $a$ and an orbital frequency
$\Omega$.  The test particle follows an initially circular orbit with
semi-major axis $a_{\rm tp}=a(1+x)$ with $x \ll 1$.  Since the
semi-major axes of the protoplanet and the test particle are close,
they rarely approach each other.  For small $x$, the angular
separation between the two bodies changes at the rate $(3/2)\Omega x$
per unit time.  Changes in the eccentricity and semi-major axis of the
test particle occur only when it reaches conjunction with the
protoplanet.

The natural scale for $x a$ is the Hill radius of the protoplanet,
$R_H \equiv (\mu/3)^{1/3} a$.  For interactions at impact parameters
larger than about four Hill radii, the effects of the protoplanet can
be treated as a perturbation to the Keplerian orbit of the test
particle.  These changes can be calculated analytically.  To first
order in $\mu$, the change in eccentricity is $e_k = A_k \mu x^{-2}$,
where $A_k = (8/9)[2 K_0(2/3)+K_1(2/3)] \approx 2.24$ and $K_0$ and
$K_1$ are modified Bessel functions of the second kind
\citep{GT78,PH86}.

The change in semi-major axis of the test particle can be calculated
from an integral of the motion, the Jacobi constant:
 $C_{\rm J} \equiv E - \Omega H$, where $E$ and $H$ and are
energy and angular momentum per unit mass of the test particle.  
Rewriting $C_{\rm J}$ in terms of $x$ and $e$, we find that

\begin{equation}
\label{eqJacobiConstant}
\frac{3}{4}x^2 - e^2 =~ {\rm const}.
\end{equation}

\noindent
If the encounter increases $e$, $|x|$ must also increase.
The change in $x$ resulting from a single interaction on an initially
circular orbit is

\begin{equation}
\label{eqDelA1Kick}
\Delta x = (2/3) e_k^2/x = (2/3) A_k^2 \mu^2 x^{-5}.
\end{equation}

The contributions of later conjunctions add to the eccentricity as
vectors and do not increase the magnitude of the eccentricity by
$e_k$. Because of this the semi-major axis of the test particle
generally does not evolve further than the initial change $\Delta x$.
Two alternatives are if the test particle is in resonance with the
protoplanet, or if its orbit is chaotic.  If the test particle is in
resonance, the eccentricity of the particle varies as
it librates.  Chaotic orbits occur when each excitation is strong
enough to change the angle of the next conjunction substantially; in
this case $e$ and $x$ evolve stochastically \citep{Wis80,DQT89}.

Orbits with $x$ between 2-4 $R_H/a$ can penetrate the Hill sphere and
experience large changes in $e$ and $a$.  This regime is highly
sensitive to initial conditions, so we only offer a qualitative
description.  Particles on these orbits tend to receive eccentricities
of the order the Hill eccentricity, $e_H \equiv R_H/a$, and
accordingly change their semi-major axes by $\sim R_H$.  We will call
this the ``strong-scattering regime'' of separations.  A fraction of
these trajectories collide with the protoplanet; these orbits are
responsible for proto-planetary accretion \citep{GL90,DT93}.

For $x \lesssim R_H/a$, the small torque from the protoplanet is
sufficient to cause the particle to pass through $x=0$.  The particle
then returns to its original separation on the other side of the
protoplanet's orbit.  These are the famous horseshoe orbits that are
related to the 1:1 mean-motion resonance.  The change in eccentricity
from an initially circular orbit that experiences this interaction can
be calculated analytically \citep{PH86}: $e_k = 2^{2/3} 3^{-3/2} 5
\Gamma(2/3) \mu^{1/3} {\rm exp}(-(8 \pi/9) \mu x^{-3})$, where
$\Gamma(2/3)$ is the usual Gamma function.  Since this interaction is
very slow compared to the orbital period, the eccentricity change is
exponentially small as the separation goes to zero.  As in the case of
the distant encounters, the conservation of the Jacobi constant
requires that $x$ increases as the eccentricity increases (equation
\ref{eqJacobiConstant}).  Then,

\begin{equation}
\label{eqDelAHorseshoe}
\Delta x = 2.83 \frac{\mu^{2/3}}{x}
{\rm exp}(-5.58 \mu x^{-3}).
\end{equation}

To apply these results to proto-planetary disks, we must allow
the test particle to have mass.  We now refer to both of the bodies as
protoplanets, each having mass ratios with the central object of
$\mu_1$ and $\mu_2$.  The change in their total separation after one
conjunction is given by equations \ref{eqDelA1Kick} and 
\ref{eqDelAHorseshoe} with $\mu = \mu_1 +\mu_2$.

\begin{figure}[t!]
\center
\includegraphics[angle=-90,width=\columnwidth]{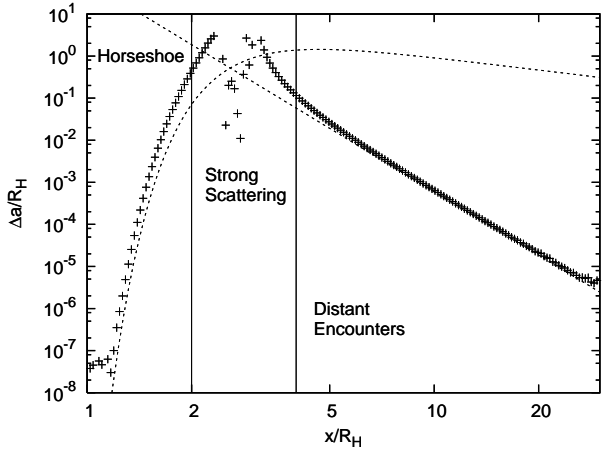}
\caption{The change in semi-major axis after a conjunction of two
bodies on initially circular orbits whose masses are smaller than that
of the star by the ratio $\mu = 3\times 10^{-9}$, plotted as a
function of the initial separation.  The points are calculated with
numerical integrations, while the dashed lines show the analytic
results, equations \ref{eqDelA1Kick} and \ref{eqDelAHorseshoe}.  At
the smallest impact parameters the bodies switch orbits; in this case
we have measured the change relative to the initial semi-major axis of
the other protoplanet.  The horizontal lines separate the regions of 
$x$ that are referred to in the text.}
\label{figDeltaA}
\end{figure}

Figure \ref{figDeltaA} plots the change in $a$ after one conjunction
of two equal mass protoplanets as measured from numerical
integrations.  All three types of interactions described above are
visible in the appropriate regime of $x$.  Each point corresponds to a
single integration of two bodies on initially circular orbits
separated by $x$.  For the horseshoe-type interactions, each
protoplanet moves a distance almost equal to $x$; we only plot the
change in separation: $\Delta a_{\rm H.S.} = |\Delta a| - |x| a$.  The
regimes of the three types of interactions are marked in the figure.
The dashed line in the low $x$ regime plots the analytic expression
calculated from equation \ref{eqDelAHorseshoe}.  The separations that
are the most strongly scattered lie between $2 - 4 R_H$, surrounding
the impact parameters for which collisions occur.  For larger
separations the numerical calculation approaches the limiting expression of
equation \ref{eqDelA1Kick}, which is plotted as another dashed line.

The sea of planetesimals modifies the dynamics of the protoplanets.
If the planetesimals have radii less than $\sim 1~{\rm km}$, their own
collisions balance the excitations caused by the protoplanets.  At the
same time, the planetesimals provide dynamical friction that damps the
eccentricities of the protoplanets.  When the typical eccentricities
of the protoplanets and the planetesimals are lower than the Hill
eccentricity of the protoplanets, this configuration is said to be
shear-dominated: the relative velocity between objects is set by the
difference in orbital frequency of nearby orbits.  In the
shear-dominated eccentricity regime, the rate of dynamical friction is
\citep{GLS04}:

\begin{equation}
\label{eqDampingRate}
- \frac{1}{e}\frac{d e}{d t} = C_d \frac{\sigma \Omega}{\rho R}\alpha^{-2}
 = \frac{1}{\tau_{\rm d}},
\end{equation}

\noindent
where $R$ and $\rho$ are the radius and density of a protoplanet,
$\sigma$ is the surface mass density in planetesimals, $\alpha$ is the
ratio $R/R_H$, and $C_d$ is a dimensionless coefficient of order
unity.  Recent studies have found values for $C_d$ between 1.2 and 6.2
(Ohtsuki et. al. 2002; Schlichting and Sari, in preparation).  For
this work, we use a value of 1.2.  For parameters characteristic of
the last stages of planet formation, $\tau_d \gg 2\pi/\Omega$.  The
interactions of the protoplanets during an encounter are unaffected by
dynamical friction and produce the change in $e$ and $a$ as described
above.  In between protoplanet conjunctions, the dynamical friction
circularizes the orbits of the protoplanets.  The next encounter that
increases $e$ further increases $x$ to conserve the Jacobi constant.
The balance between excitations and dynamical friction keeps the
eccentricities of the protoplanets bounded and small, but their
separation increases after each encounter.  This mechanism for orbital
repulsion has been previously identified by \citet{KI95}, who provide
a timescale for this process.  We alternatively derive the timescale
by treating the repulsion as a type of migration in semi-major axis.
The magnitude of the rate depends on the strength of the damping; it
is maximal if all the eccentricity is damped before the next
encounter, or $\tau_d \ll 4\pi/(3 \Omega x)$.  In this case, a
protoplanet with a mass ratio $\mu_1$ and semi-major axis $a_1$
interacting with a protoplanet with a mass ratio $\mu_2$ in the regime
of distant encounters is repelled at the rate:

\begin{equation}
\label{eqADot1Body}
\frac{1}{a_1} \frac{d a_1}{d t} = \frac{A_k^2}{2\pi} \mu_2 (\mu_1+\mu_2) x^{-4}
\Omega.
\end{equation}

\noindent
For protoplanets in the horseshoe regime, the repulsion of each
interaction is given by equation \ref{eqDelAHorseshoe}.  These
encounters increase the separation at an exponentially slower rate of:

\begin{equation}
\label{eqADotHorseshoe}
\frac{1}{a_1} \frac{d a_1}{d t} = 0.67
\mu_2 (\mu_1+\mu_2)^{-2/3}
{\rm exp}(-5.58 (\mu_1+\mu_2) x^{-3})\Omega.
\end{equation}

If instead $\tau_d \gg 4\pi/(3 \Omega x)$, the eccentricity of the
protoplanet is not completely damped away before the next conjunction
restores the protoplanet to $e\sim e_k$.  The rate at which the
separation increases is then related to the rate of dynamical
friction, $\dot a \propto e_k \dot e/x$.  Qualitatively, this rate is
slower than those of equations \ref{eqADot1Body} and
\ref{eqADotHorseshoe} by $(\tau_d \Omega x)^{-1}$.  We focus on the
maximally damped case where $\tau_d \ll 4\pi/(3 \Omega x)$.

\section{The Damped N-body Problem}
\label{secDampedNBody}

Having characterized the interactions between pairs of protoplanets,
we next examine a disk of protoplanets with surface mass density
$\Sigma$.  Each pair of protoplanets interacts according to their
separations as described in section \ref{secDamped3Body}.  If the
typical spacing is of order $R_H$, the closest encounters between
protoplanets causes changes in semi-major axes of about $R_H$ and
eccentricity excitations to $e_H$.  The strong scatterings may also
cause the two protoplanets to collide.  If the planetesimals are
shear-dominated and their mass is greater than the mass in
protoplanets, the eccentricities of the protoplanets are held
significantly below $e_H$ by dynamical friction \citep{GLS04}, and 
the distribution of their eccentricities can
be calculated analytically \citep{CS06,CSS07}.  If the scatterings and
collisions rearrange the disk such that there are no protoplanets with
separations of about $2-4 R_H$, the evolution is subsequently given by only
the gentle pushing of distant interactions \citep{KI95}.
However, there is another channel besides collisions through which the
protoplanets may achieve stability: achieving a semi-major axis very
near that of another protoplanet.

A large spacing between two protoplanets ensures they will not
strongly-scatter each other.  However, a very small difference in
semi-major axis can also provide this safety (see figure
\ref{figDeltaA} and equation \ref{eqADotHorseshoe}).  Protoplanets
separated by less than $2 R_H$ provide torques on each other during an
encounter that switch the order of their semi-major axis and reverse
their relative angular motion before they can get very close.  Their
mutual interactions are also very rare, since their relative orbital
frequency is proportional to their separation.  Protoplanets close to
co-rotation are almost invisible to each other, however these
protoplanets experience the same $\dot a/a$ from the farther
protoplanets as given by equation \ref{eqADot1Body}.  We call the
group of the protoplanets with almost the same semi-major axis a
``co-orbital group'' and use the label $N$ to refer to the number of
protoplanets it contains.  The protoplanets within a single group can
have any mass, although for simplicity in the following discussion we
assume equal masses of each.

Different co-orbital groups repel each other at the rate
of equation \ref{eqADot1Body}.  For equally spaced rows 
of the same number of equal mass protoplanets, the 
migration caused by interior groups in the disk exactly cancels the migration 
caused by the exterior groups.  We say that the protoplanets
in this configuration are separated by their ``equilibrium spacing.''
We define a quantity, $y$, to designate the distance between a single 
protoplanet and the position where it would be in equilibrium with the 
interior and exterior groups.  The near cancellation of the exterior and 
interior repulsions decreases $y$, pushing displaced protoplanets 
towards their equilibrium spacing.  The 
migration rate of a single protoplanet near the equilibrium
spacing of its group be calculated by expanding 
equation \ref{eqADot1Body} to first order in $y$ and taking the difference 
between interior and exterior contributions:

\begin{equation}
\label{eqYdot}
\frac{1}{y} \frac{d y}{d t} \approx \frac{a}{y} \sum_{i=1}^{\infty} 8 N 
%\left(
%\frac{\dot a(i x)}{a} \right)_{i x} \frac{y}{x~a} \approx
\frac{\dot a}{a} \frac{y}{i x~a} \approx
131 N \left(\frac{x~a}{R_H}\right)^{-5} e_H \Omega,
\end{equation}

\noindent
where we assume that the other co-orbital groups in the disk are
regularly spaced by $\Delta a = x~a$ and contain $N$ protoplanets of
a single mass ratio. Each term in the summation represents a pair of 
neighboring groups for which $\dot a$ is evaluated at 
the unitless separation $i x$.  
Since the repulsion rate is a sharp function of the
separation, the nearest neighbors dominate.  The coefficient in
equation \ref{eqYdot} takes a value of 121 when only the closest
neighbors are included ($i=1$ only).  Including an infinite number 
of neighbors increases the coefficent by a factor of 
$1+2^{-5}+3^{-5}+ ...$, only about 8 percent.

The dynamics above describe an oligarchic proto-planetary disk as a
collection of co-orbital groups each separated by several Hill radii.  
It is necessary though to constrain such parameters as the
typical spacing between stable orbits and the relative population of
co-orbital systems.  To determine these quantities we perform full
numerical integrations.  Given a set of initial
conditions in the strong-scattering regime, what is the configuration
of the protoplanets when they reach a stable state?

We have simulated an annulus containing twenty protoplanets, each with
a mass ratio of $\mu=1.5\times 10^{-9}$ to the central star.  The
protoplanets start on circular orbits spaced uniformly in semi-major
axis.  We dissipate the eccentricities of the protoplanets on a
timescale of 80 orbits; for parameters in the terrestrial region of
the Solar System and using $C_d = 1.2$, this corresponds to a
planetesimal mass surface density of about 8 ${\rm g~cm^{-2}}$.  We
allow the protoplanets to collide with each other setting $\alpha^{-1}
= 227$; this corresponds to a density of $5~{\rm g~cm^{-3}}$.

We examine two initial compact separations: 1.0 $R_H$ (set A) and 2.5
$R_H$ (set B).  For each initial separation we run 1000 simulations
starting from different randomly chosen initial phases.  After
$6\times 10^3$ orbital periods the orbits of the protoplanets have
stabilized and we stop the simulations.  To determine the configuration
of the protoplanets, we write an ordered list of the 
semi-major axis of the protoplanets in each simulation.  We then
measure the separation between each adjacent pair of protoplanets
(defined as a positive quantity).
If the semi-major axes of two or more protoplanets are within 2 $R_H$,
we assume they are part of the same co-orbital group.   
The average semi-major axis is calculated for each group.
The distance of each member of a group from the average 
semi-major axis we call the ``intra-group separation.''  These values 
can be either positive or negative and, for the co-orbital 
scenarios we are expecting, are typically smaller than $1 R_H$.

When one 
protoplanet is more than 2 $R_H$ from the next protoplanet, we assume that
the next protoplanet is either alone or belongs to the next co-orbital group.
The spacing between the average semi-major axis of one group and 
the semi-major axis of the next protoplanet or co-orbital group
we call the ``inter-group spacing.''  These separations are
by definition positive.

Finally we create a histogram of both the intra-group separations 
and the inter-group separations of all the simulations in the set. 
For
reference, the initial configuration of the simulations of set B 
contains no co-orbital groups.  The resulting histogram would depict
no intra-group separations, and have only one non-zero bin representing 
the inter-group separations of $x=2.5 R_H$.

\begin{figure}[t!]
\center
\includegraphics[angle=-90,width=\columnwidth]{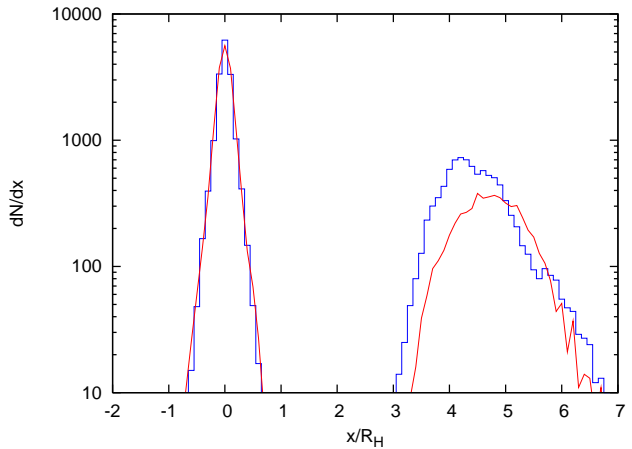}
\caption{Histogram of the intra-group and inter-group 
separations between protoplanets in
two sets of numerical simulations.  Each simulation integrates 20
protoplanets with mass ratios of $3 \times 10^{-9}$ compared to the
central mass.  They begin on circular orbits with uniform separations
in semi-major axis; each set of simulations consists of 1000
integrations with random initial phases.  The eccentricities of the
protoplanets are damped with a timescale of 80 orbits.  The smooth
line (red) represents the simulations of set A, with an initial spacing of
1.0 $R_H$, and the stepped line (blue) shows simulations of set B,
which have an initial spacing of 2.5 $R_H$.}
\label{figHistogram}
\end{figure}

Figure \ref{figHistogram} shows the histograms of the final spacings
of the two sets of simulations.  The spacings in set A are shown in
the smooth line (red), and those of set B are shown in the stepped
line (blue).  The initial closely-spaced configurations did not
survive.  The distributions plotted in figure \ref{figHistogram}
reveal that none of the spacings between neighboring protoplanets are
in the strong scattering regime, since it is unstable.  This 
validates the arbitrary choice of 2 $R_H$ as the boundary in the 
construction of figure 2; any choice between 
1 and 3 $R_H$ would not affect the results.

The size of the peak of
intra-group spacings shows that most of the protoplanets in the disk are
co-orbital with at least one other body.  The shape shows that the
spread of semi-major axis of each co-orbital group is small.  This is
consistent with equation \ref{eqYdot}, since endpoint of these
simulations is late enough to allow significant co-orbital shrinking.
The second peak in figure \ref{figHistogram} represents the 
inter-group separation.  The median inter-group separation in the two sets
are $4.8 R_H$ and $4.4 R_H$.  This is much less than the $10 R_H$
usually assumed for the spacing between protoplanets in oligarchic
planet formation \citep{KI98,KI02,Thommes03,W05}.

Figure \ref{figHistogram} motivates a description of the final
configuration of each simulation as containing a certain number of
co-orbital groups that are separated from each other by $4-5 R_H$.
Each of these co-orbital groups is further described by its occupancy
number $N$.  For the simulations of set A, the average occupancy
$\avgn = 2.8$, and for set B, $\avgn = 1.8$.  Since the simulated
annulus is small, the co-orbital groups that form near the edge are
underpopulated compared to the rest of the disk.  For the half
of the co-orbital groups with semi-major axes closest to the 
center of the annulus, $\avgn$ is higher: $\avgn
= 3.5$ for set A and $\avgn = 2.0$ for set B.

\section{Oligarchic Planet Formation} 
\label{secOligarchy}

The simulations of section \ref{secDampedNBody} demonstrate the
transition from a disordered swarm of protoplanets to an orderly
configuration of co-orbital rows each containing several protoplanets.
The slow accretion of planetesimals onto the protoplanets causes an
initially stable configuration to become unstable.  The protoplanets
stabilize by reaching a new configuration with a different average
number of co-orbital bodies.  To demonstrate this process
we simulate a disk of protoplanets and allow accretion of the
planetesimals.

We use initial conditions similar to the current picture of a disk
with no co-orbital protoplanets, placing twenty protoplanets with mass
ratios $\mu = 3 \times 10^{-9}$ on circular orbits spaced by $5 R_H$.
This spacing is the maximum impact parameter at which a protoplanet
can accrete a planetesimal \citep{GBCV91} and a typical stable spacing
between oligarchic zones (figure \ref{figHistogram}).  For the
terrestrial region around a solar-mass star, this mass ratio
corresponds to protoplanets of mass $6\times 10^{24}~{\rm g}$, far
below the final expected protoplanet mass (see section
\ref{secIsolation}).  Our initial configuration has no co-orbital
systems.  We include a mass growth term in the integration to
represent the accretion of planetesimals onto the protoplanets in the
regime where the eccentricity of the planetesimals $e_p$ obeys
$\alpha^{1/2} e_H < e_p < e_H$ \citep{DT93}:

\begin{equation}
\label{eqMdot}
\frac{1}{M} \frac{d M}{d t} = 2.4 \frac{\sigma \Omega}{\rho R}
\frac{1}{\alpha} \frac{e_H}{e_p}.
\end{equation}

\begin{figure}[t!]
\center
\includegraphics[width=\columnwidth]{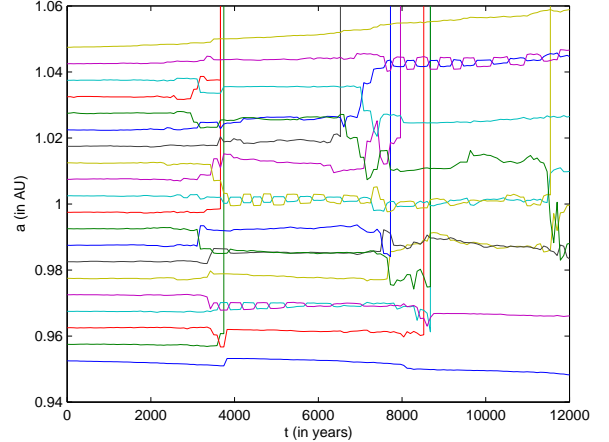}
\caption{ Semi-major axes of the protoplanets vs time in a simulation
  of oligarchic growth around a solar mass star.  The initial mass of
  each protoplanet is $6 \times 10^{24}~{\rm g}$ and each is spaced 5
  $R_H$ from its nearest neighbor.  The planetesimals have a surface
  density of $10~{\rm g~cm^{-2}}$ and an eccentricity $e_p=5 \times
  10^{-4}$.  These parameters correspond to a damping timescale of 80
  years and a growth timescale of 4800 years.  The sharp vertical
  lines indicate a collision between two bodies; the resulting
  protoplanet has the sum of the masses and a velocity chosen to
  conserve the linear momentum of the parent bodies.}
\label{figOligSim}
\end{figure}

\noindent
Protoplanet-protoplanet collisions are allowed.  For simplicity we
assume the planetesimal disk does not evolve in response to the
protoplanets.  Eccentricity damping of the protoplanets from dynamical
friction of the planetesimals is included.  The damping timescale, 80
orbits, and growth timescale, 4800 orbits, correspond to a
planetesimal surface density of $ 10~{\rm g~cm^{-2}}$ and a typical
planetesimal eccentricity of $e_p=5 \times 10^{-4}$.  We have again
used the value $C_d = 1.2$.  These parameters imply a planetesimal
radius of $\sim 100$ m, assuming that the planetesimal stirring by the
protoplanets is balanced by physical collisions.  Each protoplanet has
a density of $5~{\rm g~cm^{-3}}$.  The annulus of bodies is centered
at 1 AU.  We simulate 1000 systems, each beginning with different
randomly chosen orbital phases.  Figure \ref{figOligSim} shows the
evolution of the semi-major axis of the protoplanets in one of the
simulations as a function of time; other simulations behave similarly.

If there were no accretion, the protoplanets would preserve their
original spacing indefinitely, aside from a slow spreading at the
edges of the annulus.  However, the spacing in units of 
Hill radii decreases as the protoplanets grow.  Eventually their
interactions become strong enough to cause collisions and large
scatterings.  This epoch of reconfiguration occurs after a time of
approximately $4000$ orbits in the simulation plotted in figure
\ref{figOligSim}.  At this point the mass of protoplanets has
increased by roughly a factor of 2.3, meaning the spacing in units of
Hill radii has decreased by a factor of 1.3.  We would expect the
chaotic reconfiguration to restore the typical spacing to about $5
R_H$ by reducing the number of oligarchic zones.  The figure, in fact,
shows 13 zones after the first reconfiguration, compared to 20 before.
Three protoplanets have collided, and four have formed co-orbital groups
of $N=2$. The co-orbital pairs are visibly tightened over the
timescale predicted by equation \ref{eqYdot}, which for the parameters
of this simulation is about $\Delta t \approx 3 \times 10^3$ years.  The
configuration is then stable until the growth of the bodies again
lowers their separation into the strong-scattering regime at a time of
$1.1\times 10^4$ years.

The other realizations of this simulation show similar results.  We
find an average co-orbital population of $\avgn = 1.2$ in the middle
of the annulus after the first reconfiguration.  This value is lower
than those found in section \ref{secDampedNBody} because the
protoplanets begin to strongly-scatter each other when they are just
closer than the stable spacing.  Only a few protoplanets can collide
or join a co-orbital group before the disk becomes stable again.  As
described in the paradigm of \citet{KI95}, a realistic proto-planetary
disk in the oligarchic phases experiences many such epochs of
instability as the oligarchs grow to their final sizes.

\section{The Equilibrium Co-orbital Number}
\label{secNEq}

As the protoplanets evolve, they experience many epochs of
reconfiguration that change the typical co-orbital number.  The
examples given in previous sections of this work show the result of a
single reconfiguration.  Our choices of initial conditions with the
initial co-orbital number $\avgn_i=1$ have resulted in a higher final
co-orbital number $\avgn_f$.  If instead, $\avgn_i$ is very high, the
final co-orbital number must decrease.  As the disk evolves,
$\avgn$ is driven to an equilibrium value where each reconfiguration
leaves $\avgn$ unchanged.  This value, $\avgn_{\rm eq}$, is the number
that is physically relevant to the proto-planetary disk.

We use a series of simulations to determine $\avgn_{\rm eq}$ at a
fixed value of $\Sigma$ and $\sigma$.  Each individual simulation
contains forty co-orbital groups separated by $4~R_H$.  This spacing
ensures each simulation experiences a chaotic reconfiguration.  The
number of oligarchs in each group is chosen randomly to achieve the
desired $\avgn_i$.  All oligarchs begin with $e=e_H$ and $i=i_H$ to
avoid the maximal collision rate that occurs if $e<\alpha^{1/2}e_H$
\citep{GLS04}.  The initial orbital phase, longitude of periapse, and
line of nodes are chosen randomly.  We set a lower limit to the
allowed inclination to prevent it from being damped to unreasonably
small values.  The results of the simulations are insensitive to the
value of this limit if it is smaller than $i_H$; we choose 
$10^{-3} ~ i_H$.

We include an additional force in the simulations to prevent the
initial annulus from increasing in width.  This
extra force pushes the semi-major axis of a protoplanet back into the
annulus at a specified timescale.  We choose this timescale to be
longer than the typical time between encounters,
$(\Omega x)^{-1}$, so that multiple protoplanets
are not pushed to the boundary of the annulus without having the
chance to encounter a protoplanet a few Hill radii away.  Collisions
between protoplanets are allowed, but the 
protoplanets are not allowed to accrete the planetesimals.  
Each simulation is stopped when
there has not been a close encounter for $1.6 \times 10^4$ orbits.
Inspection of the simulation results reveals that this stopping
criteria is sufficient for the disk to have reached an oligarchic
state.  We measure the final semi-major axes of the protoplanets to
determine $N$ for each co-orbital group.  For each set of parameters
($\Sigma$, $\sigma$, and $\avgn_i$) we perform 100 simulations.

\begin{figure}[t!]
\center
\includegraphics[angle=-90,width=\columnwidth]{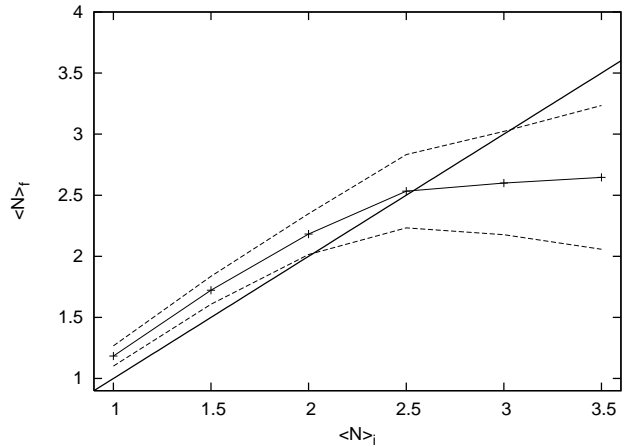}
\caption{ The final $\langle N \rangle$ of simulations against the
  initial $\langle N \rangle$ for $\Sigma=0.9~{\rm g~cm^{-2}}$ and
  $\sigma = 9.1~{\rm g~cm^{-2}}$.  For each value of $\avgn_i$ the
  mass of each protoplanet is adjusted to keep $\Sigma$ constant.  The
  dashed lines denote the average value plus and minus one standard
  deviation of the measurements.  The solid line illustrates where
  $\avgn_i=\avgn_f$.}
\label{figNivsNf}
\end{figure}

The numerical values we have chosen for these simulations reflect
planet formation in the terrestrial region.  We center the annulus of
the simulations at 1 AU.  We adopt the minimum mass solar nebula for
total mass of solids in the annulus, $\Sigma + \sigma = 10~{\rm
  g~cm^{-2}}$ \citep{Hayashi81}, and keep this value fixed throughout
all the simulations.  Figure \ref{figNivsNf} plots the results of
simulations for $\Sigma/\sigma=1/10$.  The points connected by the
solid line show the average $\avgn_f$ of each set of simulations,
while the dashed lines show the average value plus and minus 
one standard deviation of those
measurements.  For reference, we plot another solid line corresponding
to $\avgn_i = \avgn_f$.  The points at low $\avgn_i$ show a similarity
to the results of the simulations of sections \ref{secDampedNBody} and
\ref{secOligarchy}: stability is reached by increasing the number of
oligarchs in each co-orbital groups.  Once $\avgn_i$ is too high, the
chaotic reconfiguration results in an oligarchy with lower $\avgn$.
Figure \ref{figNivsNf} depicts a feedback cycle that drives
$\avgn$ towards an equilibrium value that remains unchanged by a
reconfiguration.  For $\Sigma/\sigma=1/10$, we
find $\avgn_{\rm eq} \approx 2.5$.  The intersection of the dotted 
lines with $\avgn_i=\avgn_f$ yields the one standard deviation range 
of $\avgn_{\rm eq}$, $2-3.2$.

\begin{figure}[t!]
\center
\includegraphics[angle=-90,width=\columnwidth]{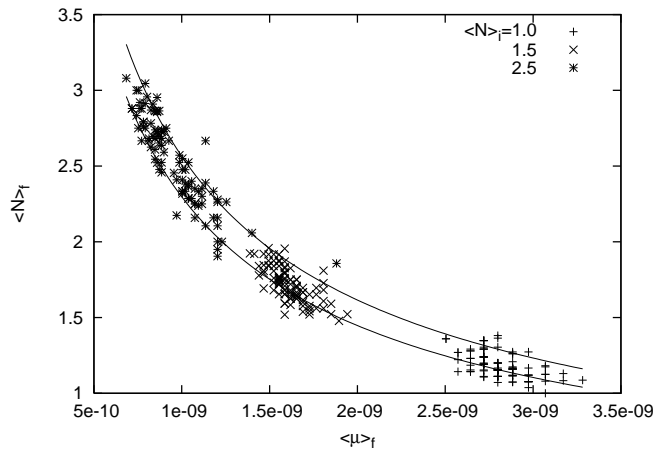}
\caption{ The final average mass ratio, $\langle \mu \rangle$, 
of the protoplanets plotted against the final $\avgn$ for 
ratio of surface densities of $\Sigma/\sigma = 1/10$. Each symbol 
corresponds to a value of $\avgn_i$.  The solid lines plot lines of 
constant $\Sigma$ for values of $\langle x \rangle$ 
one standard deviation away from the best fit curve of constant $\Sigma$ 
to the simulations with $\avgn_i=2.5$.}
\label{figMuvsNf}
\end{figure}

The cause of the wide distribution of each $\avgn_f$ is evident from
figure \ref{figMuvsNf}.  In this figure we plot the values of
$\avgn_f$ against the average mass of each protoplanet in the same
simulations of $\Sigma/\sigma=1/10$.  All of the points lie near a
single line of $\avgn_f \propto \avgm^{-2/3}$.  This relation is
derived from the definition $\Sigma=N m_p/(2 \pi \Delta a a)$.  We
find the relation

\begin{equation}
\label{eqAvgX}
\avgn = \frac{2 \pi a^2 \Sigma}{3^{1/3} M_{\odot}} \avgx
\langle \mu \rangle^{-2/3},
\end{equation}

\noindent
where we have defined $x_H$ to be dimensionless and equal to $\Delta
a/R_H$.  While the points in figure \ref{figMuvsNf} generally follow
the function given by equation \ref{eqAvgX}, there is significant
scatter.  We interpret this variation as a distribution of the average
spacing between rows, $\avgx_f$.  For the $\avgn_i=2.5$ simulations,
we measure an average $\avgx_f=5.4$, with a standard deviation of 0.2.
The solid lines in figure \ref{figMuvsNf} correspond to the lower and
upper bounds of $\avgx_f$ given by one standard deviation from the mean.
This reaffirms our earlier conclusion that the spacing between rows is
an order unity number of Hill radii of an average size body. 

\begin{figure}[t!]
\center
\includegraphics[angle=-90,width=\columnwidth]{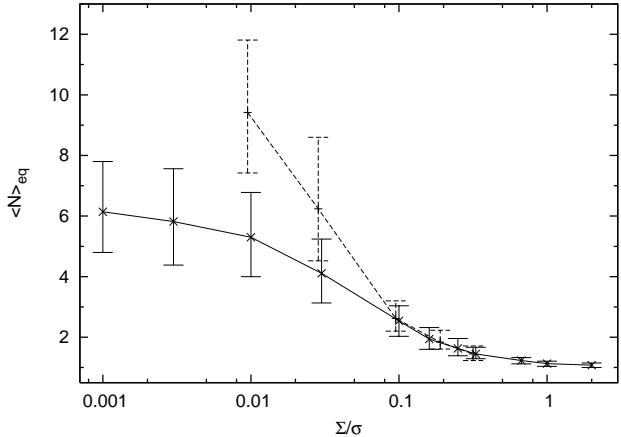}
\caption{ The equilibrium average co-orbital number $\avgn_{\rm eq}$ plotted 
against the surface mass density ratio of protoplanets to planetesimals,
$\Sigma/\sigma$.  The error bars represent the standard deviation 
of $\avgn_{\rm eq}$ as defined in the text.  The solid and dashed
points correspond to simulations at 1 AU and 25 AU respectively.
The dashed points are offset by 5 \% in $\Sigma/\sigma$ 
to distinguish them from the solid points.}
\label{figNeqvsSigma}
\end{figure}

The ratio of $\Sigma/\sigma$ increases as the oligarchs accrete the
planetesimals.  To demonstrate the evolution of $\avgn_{\rm eq}$ and
$\avgx_{\rm eq}$, we performed more simulations with values of
$\Sigma/\sigma$ in the range 0.001-2.  At each value we examine a range of
$\avgn_i$ to determine $\avgn_{\rm eq}$.  We plot the resulting values
in figure \ref{figNeqvsSigma}.  The error bars on the points show
where one standard deviation above and below $\avgn_f$ is equal to
$\avgn_i$.  As the disk evolves and $\Sigma/\sigma$ approaches unity,
$\avgn_{\rm eq}$ decreases.  For high values of $\Sigma/\sigma$, the
equilibrium co-orbital number asymptotes towards its minimum value by
definition, 1.

\begin{figure}[t!]
\center
\includegraphics[angle=-90,width=\columnwidth]{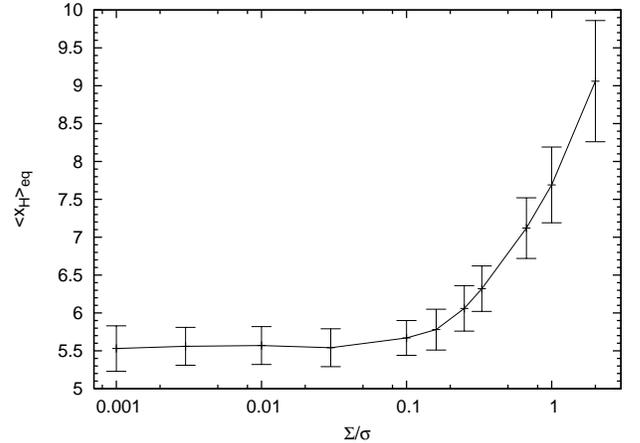}
\caption{ The equilibrium average spacing between co-orbital groups,
$\avgx_{\rm eq}$ in for simulations with $\avgn_i=\avgn_{\rm eq}$
plotted against the surface mass density ratio $\Sigma/\sigma$.
The error bars reflect the standard deviation of the measurements
of $\avgx$ of each simulation.
}
\label{figXeqvsSigma}
\end{figure}

For the simulations with $\avgn_{\rm eq}$, we also measure the average
spacing between co-orbital groups directly.  The average spacing in
units of the Hill radii of the average mass protoplanet, $\avgx_{\rm
  eq}$ is plotted against $\avgn_{\rm eq}$ in figure
\ref{figXeqvsSigma}.  Early in the disk, when $\Sigma/\sigma$ is very
small, $\avgx_{\rm eq}$ is approximately constant at a value of 5.5.
The average spacing grows however as $\Sigma/\sigma$ approaches unity.

\begin{figure}[t!]
\center
\includegraphics[angle=-90,width=\columnwidth]{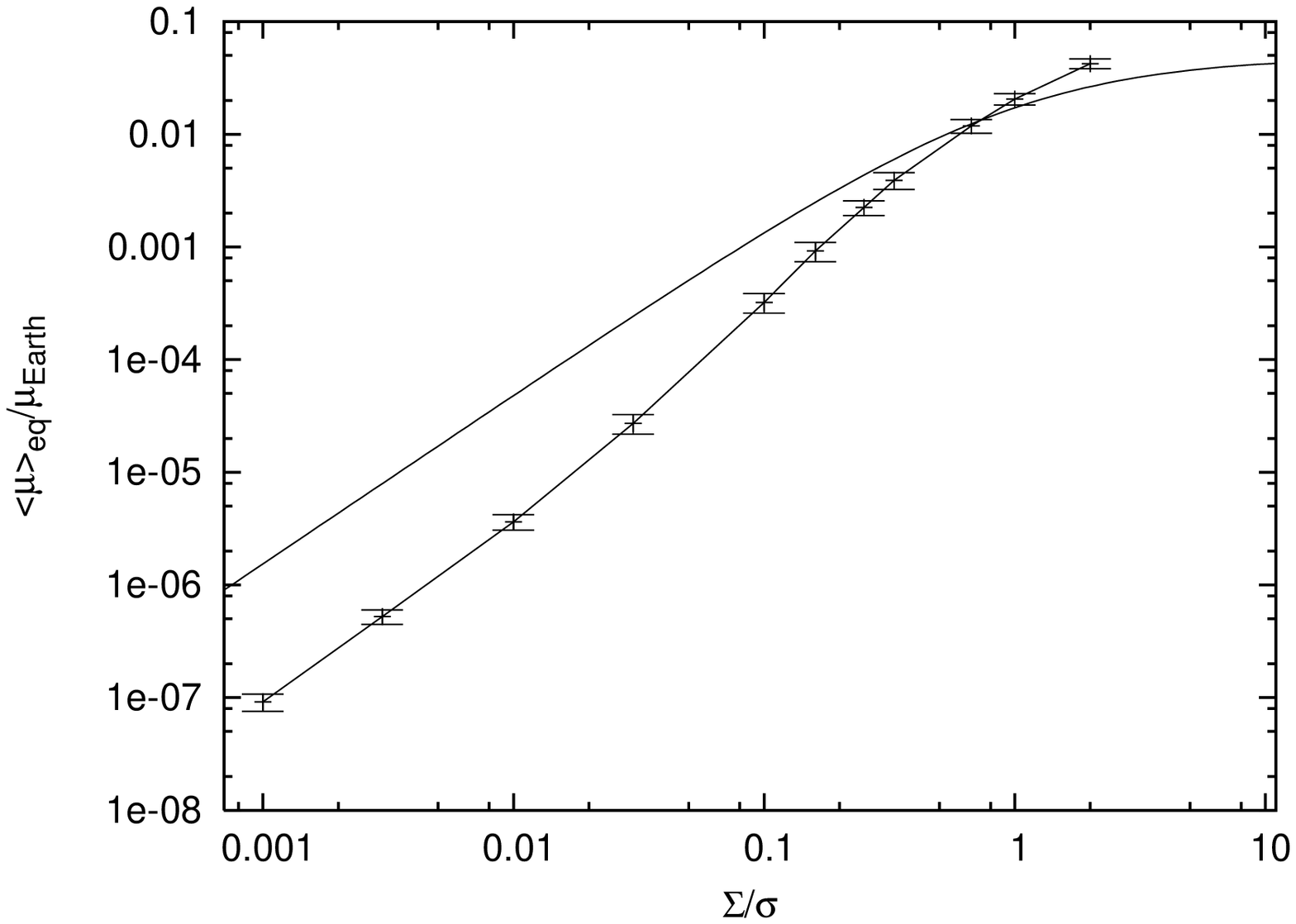}
\caption{ The average mass of the protoplanets in an equilibrium
oligarchy as a function of the surface mass density ratio
$\Sigma/\sigma$ at $a=1$ AU.  The error bars are the standard deviation in
average mass of the simulations for $\Sigma/\sigma$ and
$\avgn_i=\avgn_{\rm eq}$.  The solid line plots the average
protoplanet mass given by an $\avgn=1$ and $\avgx=5$ oligarchy
commonly assumed in the literature, described by equation \ref{eqAvgX}.
}
\label{figMpeqvsSigma}
\end{figure}

Figure \ref{figMuvsNf} shows that all oligarchies of a fixed $\Sigma$
exhibit similar average spacings $\avgx$.  The points from simulations
of different $\avgn_i$ confirm that a broad range of $\avgn$ and
$\avgm$ can be achieved, with the relation between $\avgn$ and $\avgm$
given by equation \ref{eqAvgX}.  By finding the equilibrium $\avgn$
reached by the disk after many configurations, we also fix the average
mass of the protoplanet, denoted $\avgm_{\rm eq}$.  We plot
$\avgm_{\rm eq}/\mu_{\rm Earth}$ as a function of $\Sigma/\sigma$ at
$a=1$ AU in figure \ref{figMpeqvsSigma}, where $\mu_{\rm Earth}$ is
the mass ratio of the Earth to the Sun.  The error bars show the
standard deviation of $\avgm$ for the simulations with
$\avgn_i=\avgn_{\rm eq}$.

For comparison, we also plot $\avgm$ as given by equation \ref{eqAvgX}
for a constant $\avgn_i=1$ and $\avgx=5$.  These parameters reflect
the typical oligarchic picture with no co-orbital oligarchs and a
fixed spacing in Hill units \citep{Lis87,KI95,GLSfinalstages}.  At low
$\Sigma/\sigma$, the solid line over-estimates the protoplanet mass by 
over an order of magnitude.
This is a result of large $\avgn_{\rm eq}$, which allows the disk mass
to be distributed into several smaller bodies instead of a single
protoplanet in each oligarchic zone.  For $\Sigma/\sigma$ greater than
about 0.5, the lines cross, and the simple picture is an underestimate
of $\avgm_{\rm eq}$.  Although $\avgn_{\rm eq}$ is close to one for
these disks, $\avgx_{\rm eq}$ grows, increasing the relative amount of
the total disk mass that has been accreted into each protoplanet.

%another batch of simulations using
%surface densities such that $\Sigma/\sigma=1/4$.  We find in this case
%that $\avgn_{\rm eq} \approx 1.6$, with the standard deviations
%allowing a range of $1.4 - 1.9$.  The average spacing between rows for
%this $\avgn$ is measured to be $\avgx_{\rm eq}=5.8$.  In general, as
%oligarchy proceeds, $\avgn_{\rm eq}$ decreases and $\avgx_{\rm eq}$.

We performed the same calculations for several sets of simulations
with the annulus of protoplanets centered at 25 AU.  The values of
$\avgn_{\rm eq}$ we find for these simulations are plotted as the
dashed line in figure \ref{figNeqvsSigma}.  For $\Sigma/\sigma <
0.1$, the co-orbital groups tend to contain more oligarchs at 
25 AU than at 1 AU, but the spacing between rows is still 
$\avgx_{\rm eq} \approx 5.5$.  For larger $\Sigma/\sigma$, the  
distance of the protoplanets from the star matters less.

\section{Isolation}
\label{secIsolation}

Oligarchic growth ends when the protoplanets have accreted most of the
mass in their feeding zones and the remaining planetesimals can no
longer damp the eccentricities of the protoplanets.  The
eccentricities of the protoplanets then grow unchecked; this is
known as the ``isolation'' phase.  The mass of a protoplanet at this
point is referred to as the ``isolation mass,'' and can be found 
from equation \ref{eqAvgX}:

\begin{equation}
\label{eqMiso}
\frac{M_{\rm iso}}{M_{\rm star}} = \frac{1}{3^{1/2}}
\left[
\left(\frac{\Sigma/\sigma}{\Sigma/\sigma+1}\right) 
\frac{M_{\rm disk}}{M_{\rm star}} 
\frac{\avgx}{\avgn} \right]^{3/2}.
\end{equation}

\noindent
The literature typically assumes that at isolation all of the mass is 
in protoplanets.  This is equivalent to the limit of $\Sigma/\sigma \gg 1$.

The results of section \ref{secNEq} show that oligarchy at a fixed
semi-major axis is uniquely described by $\Sigma/\sigma$.  For the
terrestrial region then, $M_{\rm iso}$ is given by the parameters we
calculate in section \ref{secNEq}, and is plotted as a function of
$\Sigma/\sigma$ in figure \ref{figMpeqvsSigma}.

The exact ratio of mass in protoplanets to that in planetesimals that
allows the onset of this instability in the terrestrial region is not
known; simulations suggest that in the outer solar system this
fraction $\Sigma/\sigma \approx 10$ \citep{FC07}.  It is not
straightforward to determine the value of $\Sigma/\sigma$ for which
isolation occurs.  In many of our simulations, the eccentricities of 
the protoplanets rise above $e_H$, yet an equilibrium is
eventually reached.  We postpone a detailed investigation of the
dynamics of the isolation phase for a later work.  For any value of
$\Sigma/\sigma$ at isolation however, the properties of the oligarchy
at this stage can be read from figures
\ref{figNeqvsSigma},\ref{figXeqvsSigma}, and \ref{figMpeqvsSigma}.

%In our new picture where many oligarchs find safety through
%co-rotation, the mass contained in one feeding zone must be divided
%among the $N$ oligarchs in that co-orbital group.  The characteristic
%mass of the oligarchs at this point, $M_{\rm iso}$, is related to the
%total mass of the disk for a given spacing between co-orbital groups
%and a typical occupancy number $\avgn$ \citep{Lis87,GLSfinalstages},

%\begin{equation}
%\label{eqMiso}
%\frac{M_{\rm iso}}{M_{\rm star}}=6.5 \left(\frac{f}{\avgn}
%\frac{M_{\rm disk}}{M_{\rm star}}\right)^{3/2},
%\end{equation}

%\noindent
%where $f$ is the ratio of protoplanet to planetesimal mass at the time
%of the eccentricity instability, and $M_{\rm disk} = 2\pi( \Sigma
%+\sigma) a^2$ is the mass contained in an interval of width $a$.  We
%have used a spacing between co-orbital groups of $5 R_H$ as we found
%in our simulations.

The fate of the protoplanets after isolation depends on their distance
from the star.  In the outer parts of the solar system, the nascent
ice giants are excited to high eccentricities and may be ejected from
the system entirely \citep{GLSfinalstages,FC07,LM07}.  Their lower
rate of collisions also likely increases their equilibrium co-orbital
number for a fixed $\Sigma/\sigma$ relative to this work performed in
the terrestrial region.  In contrast to giant impacts, ejections do
not change the mass of individual protoplanets, so they must
reach their full planetary mass as oligarchs.  For an $\avgn \neq 1$
at isolation, the mass of the disk needs to be augmented
proportionally to $\avgn$ so that $\avgm_{eq}$ at isolation is equal
to the mass of an ice giant.
%If we
%assume that $f=0.5$ and $\avgn=2$, we require the Minimum Mass Solar
%Nebula \citep{Hayashi81} to be enhanced by a factor of 12 in order to
%form Neptune- and Uranus-sized isolation masses.  This problem is
%alleviated slightly if the protoplanets contain most of the disk mass
%at isolation \citep{FC07}.

The terrestrial planets tend to collide before they can be ejected, as
the escape velocity from their surfaces is smaller than the velocity
needed to unbind them from solar orbits
\citep{Cham01,GLSfinalstages,KB05}.  This process conserves the total
mass of protoplanets so $M_{\rm disk}$ is given by the Minimum Mass
Solar Nebula.  Accounting for $\avgn \neq 1$ in this case reduces the
mass of each body at isolation proportionally to $\avgn^{3/2}$.
This in turn increases the number of giant impacts necessary to
assemble the terrestrial planets.

%by a factor of 3; equation \ref{eqMiso}
%shows that $M_{\rm iso} \propto \avgn^{-3/2} \approx 3\times 10^{-3}
%M_{\Earth}$ for $f=0.5$.  Correspondingly, the number of giant impacts
%necessary to assemble the terrestrial planets is about 300.

\section{Conclusions and Discussion}
\label{secConclusions}

We have analyzed the interactions of a disk of protoplanets
experiencing dynamical friction.  Conjunctions of a pair of
protoplanets separated by more than 3 $R_H$ increase the separation of
that pair.  The repulsions from internal protoplanets
cancel those from external protoplanets at a
specific equilibrium semi-major axis.  Several bodies can inhabit this
semi-major axis on horseshoe-like orbits.  We have shown through
numerical simulations that these co-orbital systems do form and
survive.  We expect the oligarchic phase of planet formation to
proceed with a substantial population of co-orbital protoplanets.  We
present an empirical relation between the ratio of masses in
protoplanets and planetesimals, $\Sigma/\sigma$, and the equilibrium
average co-orbital number $\avgn_{\rm}$ and the equilibrium average
spacing between co-orbital groups $\avgx_{\rm}$.  To form the extra
ice giants that populate the co-orbital groups in the outer solar
system, the mass of the proto-planetary disk must be enhanced by
$\avgn$ relative to the existing $N=1$ picture.  To form the
terrestrial planets requires $\avgn^{3/2}$ more giant impacts.  While
we have not calculated the critical value of $\Sigma/\sigma$ that
initiates the isolation phase, we have completely determined the
parameters of a shear-dominated oligarchy of protoplanets up to that
point.

In section \ref{secDampedNBody}, we have ignored the repulsive distant
interactions between a protoplanet and the planetesimals that cause
type I migration \citep{GT80,W86}.  The additional motion in
semi-major axis is only a mild change to the dynamics.  In a uniform
disk of planetesimals, an oligarchic configuration of protoplanets
migrates inward at an average rate specified by the typical mass of
the protoplanets.  Mass variation between the protoplanets of
different co-orbital groups causes a differential migration relative
to the migration of the entire configuration.  However, the repulsion
of the neighboring co-orbital groups counteracts the relative
migration by displacing the equilibrium position between two groups by
an amount $ \sim (\sigma/\Sigma) (R_H/a) R_H$.  Differential migration also
acts on members of a single co-orbital group, however its effects
cannot accumulate due to the horseshoe-like co-orbital motion.  The
ratio of the timescale for migration across the co-orbital group to
the interaction timescale sets a minimum safe distance from the
equilibrium separation: $y_{\rm safe}/R_H \sim \mu^{-1/6} (M_{\rm
  disk}/M_{\odot})^{1/2}$.  For typical co-orbital group, where $y
\sim R_H$, the migration is never fast enough for a protoplanet to
escape the group before the next encounter with a co-orbiting
protoplanet brings it to the other side of the nominal equilibrium
semi-major axis.

It is also possible that the disk of planetesimals is not uniform.
The accretional growth of a protoplanet may lower the surface density
of planetesimals at that semi-major axis such that the total mass is
locally conserved.  One might naively expect that the deficit of
planetesimals exactly cancels the repulsion caused by the formed protoplanet.
However, it can be seen from equation \ref{eqADot1Body} that the rate
of repulsion of a protoplanet from another protoplanet of comparable
mass is twice that of the same mass in planetesimals.  The net rates of
repulsion of the protoplanets in this scenario are reduced by a factor
of two; the dynamics are otherwise unchanged.

One important question is that of the boundary conditions of a
planet-forming disk.  The initial conditions of the simulations we
present only populate a small annulus around the central star.  We
artificially confine the bodies in this region to force the surface
mass density to remain constant.  The behavior of $\Sigma$ over a
larger region of the disk may not be similar to that of our annulus.
The presence of gas giants or previously formed planets may prevent
any wide-scale diffusion of protoplanets across the disk.  On the
other hand, the dynamics in a logarithmic interval of semi-major axis
may not be affected by the populations internal and exterior to that
region.  The behavior of protoplanets in the oligarchic phase in a 
full size proto-planetary disk is an open question.

Earlier analytical work has examined the interactions between
oligarchs that share a feeding zone \citep{GLS04}.  These authors
conclude that protoplanets in an oligarchic configuration are always
reduced to an $\avgn=1$ state.  However, we have shown that for a
shear-dominated disk, the collision rate between protoplanets is
suppressed as the protoplanets are pushed towards almost the same
semi-major axis.  The growth rate of the protoplanets of each
co-orbital group depends on the eccentricity of the planetesimals.
For $e_p<\alpha^{1/2} e_H$ the growth rate of a protoplanet scales as
$R^{-1}$.  This is called ``orderly'' growth since all of the
protoplanets approach the same size.  In the intermediate
shear-dominated regime of $\alpha^{1/2} e_H < e_p < e_H$, the growth
rate is independent of $R$.  The protoplanets then retain the relative
difference in their sizes as they grow.  For shear-dominated disks,
which are the focus of this paper, the co-orbital groups are not
disrupted by differential growth.

The spacing between co-orbital groups that we observe for most
$\Sigma/\sigma$ is smaller than the $10 R_H$ that is typically assumed
\citep{KI98,KI02,Thommes03,W05} based on the simulations by \citet{KI98}.
Their simulations are in the dispersion-dominated eccentricity regime,
where the maximum distance at which an oligarch can accrete a
planetesimal is set by the epicyclic motion of the planetesimals,
$\sim e a$.  This motion sets the width of the feeding zones; the
figures of \citet{KI98} indicate that the typical eccentricity of the
smaller bodies corresponds to a distance of $10 R_H$.
Dispersion-dominated disks with different values for protoplanet sizes
and planetesimal eccentricities should undergo oligarchy with a
different spacing.  In shear-dominated disks, we have shown that
separations of about $5 R_H$ are set by the distant encounters with
the smallest impact parameters.

The simulations of \citet{KI98} do not contain any co-orbital
groups of protoplanets; this is expected due to the small number of
protoplanets that form in their annulus and the fact that their 
eccentricities are super-Hill.  \citet{Thommes03}
examine a broad range of parameters of oligarchic growth, but the
number of planetesimals are not enough to damp the protoplanet
eccentricities sufficiently.  However, upon inspection of their figure
17 we find hints of the formation of co-orbital groups.  Also, even
though a range of separations are visible, many adjacent feeding zones
are separated by only $5 R_H$ as we are finding in our simulations.

Simulations of the oligarchic phase and the isolation epoch that
follows by \citet{FC07} include five bodies that are spaced
safely by $5 R_H$.  We would not expect the formation of co-orbital
oligarchs from an initial state of so few.  Interestingly,
\citet{LM07} use a population of ``tracer particles'' to calculate the
effects of planetesimals on their protoplanets and find a strong
tendency for these objects to cluster both in co-orbital resonances
with the protoplanets and in narrow rings between the protoplanet
orbits.  This behavior can be understood in light of our equation
\ref{eqDelA1Kick} with the dynamical friction of our simulations
replaced by the collisional damping of the tracer particles.

Simulations of moderate numbers of protoplanets with eccentricity
damping and forced semi-major axis migration were studied by
\citet{CN06}; indeed they observe many examples of the co-orbital
systems we have described.  We offer the following comparison between
their simulations and this work.  Their migration serves the same
purpose as the growth we included in the simulations of section
\ref{secOligarchy}, namely to decrease the separations between bodies
until strong interactions rearrange the system with stable spacings.
The co-orbital systems in their simulation likely form in the same way
as we have described: a chance scattering to almost the same
semi-major axis as another protoplanet.  They attribute the tightening
of their orbits to interactions with the gas disk that dissipates
their eccentricity, however, this is unlikely.  Although very close in
semi-major axis, in inertial space the co-orbital protoplanets are
separated by $\sim a$ for most of their relative orbit. Since the
tightening of each horseshoe occurs over only a few relative
orbits, it must be attributed to the encounters with the other
protoplanets, which occur more often
than the encounters between the co-orbital pairs.  

Cresswell and
Nelson also find that their co-orbital pairs settle all the way to
their mutual L4 and L5 Lagrange points; the systems that we describe
do not.  In our simulations a single interaction between neighbors
moves each protoplanet a distance on the order of the width of the
largest possible tadpole orbit, $\Delta a/a \sim \mu^{1/2}$.  The
objects in the simulations by Cresswell and Nelson have much larger
mass ratios with the central star and larger separations.  In their
case a single interaction is not strong enough to perturb the
protoplanets away from the tadpole-like orbits around the Lagrange
points.  We have performed several test integrations with parameters
similar to those run by Cresswell and Nelson and confirmed the formation 
of tadpole orbits.  Finally, their simulations model the end of the
planet formation and hint at the possibility of discovering extrasolar
planets in co-orbital resonances.  In a gas depleted region, we do not
expect the co-orbital systems that form during
oligarchic growth to survive the chaos following isolation.

In the terrestrial region of the solar system, geological measurements
inform our understanding of the oligarchic growth phase.  Isotopic
abundances of the Martian meteorites, in particular that of the
Hafnium (Hf) to Tungsten (W) radioactive system, depend on the
timescale for a planet to separate internally into a core and mantle.
Based on these measurements, \citet{HK06} calculate that Mars
differentiated quickly compared to the timescale of the Hf-W decay, 9
Myrs.  The oligarchic picture of equation \ref{eqAvgX} with $\avgn=1$
shows that at 1.5 AU with $\avgn=1$, and $\Sigma \sim \sigma$, $\avgm
\approx M_{\rm Mars}/M_{\odot}$; accordingly these authors infer that
Mars was fully assembled by the end of the oligarchic phase and did
not participate in the giant impacts that assembled Earth and Venus.
A co-orbital oligarchy, however, lowers the mass of each protoplanet
at isolation by a factor of $\avgn^{3/2}$.  In this picture
Mars formed through several giant impacts.  This scenario is
consistent with the isotopic data if Mars can experience several
collisions in 10 Myrs; the collisional timescales for $\avgn>1$
systems merit further investigation.

The rate and direction of the rotation of Mars, however, provide
further evidence for a history of giant impacts.  \citet{DT93}
calculate the angular momentum provided by the collision-less
accretion of planetesimals and show that, for any planetesimal
velocity dispersion, this process is insufficient to produce the
observed spins.  The moderate prograde rotation of Mars is thus
inconsistent with pure accretionary growth.  \citet{SS06} show that
the collisions of planetesimals inside the Hill sphere as they accrete
produces protoplanets that are maximally rotating, which is still
inconsistent with the current rotation of Mars.  Giant impacts later
re-distribute the spin-angular-momentum of the protoplanets but with a
prograde bias; this then implies that Mars did participate in the
giant impact phases of the terrestrial region.  Again, further studies
are necessary to characterize the timescale of the collisional period
following the isolation phase in an $\avgn > 1$ scenario.

The compositions of the planets offer more clues to their formation.
As protoplanets are built up from smaller objects in the
proto-planetary disk, their composition approaches the average of the
material from which they accrete.  Numerical simulations by
\citet{Cham01} show that the collisional assembly of protoplanets
through a $\avgn=1$ oligarchy mixes material from a wide range of
semi-major axes.  The composition of the planets then reflects some
average of all available material.  The three stable isotopes of
oxygen are thought to be initially heterogeneous across the
proto-planetary disk, and offer a measurable probe of compositional
differences between solar system bodies.  In the case of the Earth and
Mars, a small but finite difference in the ratios of these isotopes is
usually attributed to the statistical fluctuations of the mixing
process \citep{FBBW01,OPHY07}.  An $\avgn > 1$ oligarchy requires more
collisions; the same isotopic variance between Earth and Mars may
require a larger dispersion in the composition of the smallest
proto-planetary materials.  However, it is necessary to determine the
extent of spatial mixing in the $\avgn>1$ picture and to understand
the changes in composition resulting from a single giant impact
\citep{PS07} before we can estimate the primordial compositional
variations allowed by this model.

We thank Dave Stevenson for enlightening discussions.  Insightful
comments by our referee, Eiichiro Kokubo, motivated significant
improvements to this work. R.S. is a Packard Fellow and an Alfred
P. Sloan Fellow.  This work was partially supported by the European
Research Council (ERC).

%\bibliographystyle{apj}
%\bibliography{ms}

%\clearpage

\end{document}